\documentclass[pre,twocolumn,amsmath,amssymb,aps,superscriptaddress]{revtex4}
\usepackage{graphicx,color}
\usepackage{dcolumn}
\usepackage{bm}
\usepackage{textcomp}
\usepackage {diagbox}

\def\tN{\widetilde{N}}
\def\d{\delta}
\def\e{\epsilon}
\def\g{\gamma}

\def\lam{\lambda}
\def\hC{\widehat{C}}
\def\bA{\mathbb{A}}
\def\bS{\mathbb{S}}

\def\m{\mathfrak{m}}

\def\j{\mathfrak{j}}
\def\p{\partial}
\def\l{\left}
\def\r{\right}
\def\la{\langle}
\def\ra{\rangle}

\def\f{\frac}

\begin{document}

\title{Energy current correlation in solvable long-range interacting systems}

\author{Shuji Tamaki}
\affiliation{Department of Physics, Keio University, Yokohama 223-8522, Japan}

\author{Keiji Saito}
\affiliation{Department of Physics, Keio University, Yokohama 223-8522, Japan}

\date{\today}

\begin{abstract}
  We consider heat transfer in one-dimensional systems with long-range interactions. It is known that typical short-range interacting systems shows anomalous behavior in heat transport when total momentum is conserved, whereas momentum-nonconserving systems do not exhibit anomaly. In this study, we focus on the effect of long-range interaction. We propose an exactly solvable model that reduces to the so-called momentum-exchange model in the short-range interaction limit. We exactly calculate the asymptotic time-decay in the energy current correlation function, which is related to the thermal conductivity via the Green--Kubo formula. From the time-decay of the current correlation, we show three qualitatively crucial results. First, the anomalous exponent in the time-decay {\it continuously} changes as a function of the index of the long-range interaction. Second, there is a regime where the current correlation diverges as increasing the system size with fixed time, and hence the exponent of the time-decay cannot be defined. Third, even momentum-nonconserving systems can show the anomalous exponent indicating anomalous heat transport. Higher-dimensions are also considered, and we found that long-range interaction can induce anomalous exponent even in the three-dimensional systems.
\end{abstract}

\maketitle

\section{\label{sect:intro}INTRODUCTION}

In the past few decades, the study of dynamic and thermodynamic properties of long-range interacting systems has attracted considerable attention. These systems are characterized by interaction potentials $V(r)$ that decay with the power law
\begin{align}
  V(r)\propto r^{-\d}\, ,
\end{align}
where $r$ is the distance between two interacting particles. The parameter $\d$ controls the range of interaction; a smaller $\d$ means longer range of interactions. When the index $\d$ is lower than the spatial dimension, the system is called the long-range interacting system \cite{campa2009physrep,campa2014book}. In this regime, the additivity does not hold and many unusual properties appear such as negative specific heat \cite{thirring,hertel-thirring,antoni-torcini-1998,barre-mukamel-ruffo}, long-lived quasistationary state \cite{latora-rapisarda-tsallis,christodoulidi-tsallis-bountis}, anomalous diffusion \cite{antoni-torcini-1998,torcini-antoni-1999,latora-rapisarda-ruffo-1999}, and suppression of chaos \cite{latora-rapisarda-ruffo-1998,anteneodo-tsallis,torcini-antoni-1999,christodoulidi-tsallis-bountis,bagchi-tsallis}. In this study, we use the terminology {\it long-range interaction} in a wider sense, to refer to the interaction in the power-law form regardless of the exponent $\d$.

In contrast to the equilibrium properties, nonequilibrium properties such as transport have not yet been understood in such systems. In this study, we address the heat transfer in long-range interacting systems focusing on energy current fluctuations. We focus on one-dimensional systems, because there are many studies on the short-range interacting systems. In short-range interacting one-dimensional systems with total momentum conservation, the energy transport is in general anomalous, in the sense that the thermal conductivity $\kappa$ diverges as $\kappa\propto N^{\alpha}$ ($0 < \alpha \le 1$) with an increase in the system size $N$ \cite{lepri2003physrep,dhar2008,lepri2016book}. The thermal conductivity is given by the Green--Kubo formula, which is the time integral of the energy current correlation. Hence, the anomalous behavior of the diverging conductivity is directly related to the slow decay in the equilibrium current correlation in a closed system:
\begin{align}
  C (t) &:= N^{-1} \langle J_{\rm tot} (t) J_{\rm tot}  \rangle \sim t^{-\beta } \, , ~~~ 0 \le \beta < 1 \, , \label{ctintro}
\end{align}
where $J_{\rm tot}$ is the total energy current, and $\langle ...\rangle$ is the equilibrium average or microcanonical average. In addition, from the microscopic viewpoint, this slow relaxation is also related to the super-diffusive behavior in the energy diffusion\cite{jara2009,basile-olla-spohn-2010,jara2015}. If the system has onsite pinning potentials where the total momentum conservation does not hold, the above anomaly disappear, and the normal diffusion as well as normal heat transport are recovered.
\begin{table}[t]
  \begin{flushleft}
  \begin{tabular}{|c | c | c | c | c|}
    \hline   
        &~ $\d \le 2$~ & $2 < \d <  3$ & $3 < \d$ \\
    \hline
     $k_0 =  0 $  & --- & $ \beta ={(\delta -2 )/2}  $ & $\beta={1/   2 }$ \\
            & ---&  (anomalous)  & (anomalous)  \\
    \hline   
  \end{tabular}
  \\
  \begin{tabular}{|c | c | c | c | c |}
    \hline   
    & ~$\d \le 3/2$ ~& $3/2 < \d \le 5/2$& $5/2 < \d < 3$  & $3 < \d$ \\
    \hline
    $k_0 \neq  0 $    & ---      & $\beta= (2 \delta -3)/2$ & $\beta= (2 \delta -3)/2 $& $\beta={3/2}$ \\
                & --- &  (anomalous)             & (diffusive)       & (diffusive)  \\
    \hline   
  \end{tabular}
\end{flushleft}
\label{tone}
\caption{Decay rate $\beta$ for the one-dimensional long-range interacting systems. The upper table summarizes the systems with no onsite potential. The lower table summarizes the results for systems with onsite potential. Note that the exponent gradually changes depending on $\d$. Remarkably, even systems with onsite potentials can exhibit anomalous decay for the range $3/2 < \d \le 5/2$. At $\d=3$, the logarithmic correction appears in the time-decay for both cases of $k_0=0$ and $k_0\neq 0$ (See Eqs.(\ref{eq:limt-C}) and Eqs.(\ref{eq:limt-C:onsite})). In the regime $\d \le 2$ for $k_0=0$ and the regime $\d \le 3/2$ for $k_0\neq 0$, the amplitude of current correlation diverges; hence, the exponent cannot be defined.}
  \end{table}

From the above backgrounds on short-range interacting systems, we consider the effects of long-range potentials on energy fluctuations. Thus far, several numerical studies have proposed two paradigmatic models. In Refs.\cite{olivares2016,bagchi2017xy,iubini2018}, the coupled rotor model was studied. This model shows a transition from the diffusive transport to the thermal insulator, as $\d$ decreases from infinity. The critical point $\d_c$ lies between $\d=1$ and $2$; however, the explicit value depends on the temperature regime. In Refs.\cite{bagchi2017fpu,iubini2018}, the Fermi--Pasta--Ulam (FPU) model was investigated, and it was found that the transport behavior is generally anomalous, except at $\d =2$, where it exhibits ballistic behavior \cite{bagchi2017fpu}. In the presence of long-range interactions, the coupling form between the system and the reservoir is very nontrivial in the choice, since the interaction range of the coupling (i.e., long-range or short-range) may cause significant macroscopic difference. In the recently study \cite{xiong}, the reservoir effect is carefully studied, and several intriguing effects including energy diffusion behavior are discussed. In studies on the FPU model, long-range interactions can be added in several ways, e.g., long-range quadratic potential terms\cite{miloshevich}, long-range quartic potential terms \cite{bagchi-tsallis,bagchi2017fpu}, and a combination of both \cite{christodoulidi-tsallis-bountis,iubini2018,xiong}. From these studies, many intriguing transport properties have been numerically indicated. However, note that, in general, it is very difficult to obtain clear results through numerical calculations, because finite-size effect is very significant, especially in long-range systems \cite{iubini2018}. Owing to this difficulty, we require clear-cut results with a solvable model for an in-depth understanding of the long-range effect.

In this paper, we propose an analytically solvable model, which mimics the FPU dynamics with the long-range quadratic potential. The dynamics of the model consists of the deterministic Hamiltonian dynamics of harmonic interactions and stochastic perturbation exchanging momentums of nearest neighbors. The Hamiltonian for the deterministic dynamics is given in Eqs.(\ref{eq:H}) and (\ref{eq:H1}). For $\d=\infty$, without onsite pinning potential, this model is equivalent to the so-called momentum exchange (ME) model \cite{basile2006prl,basile2009cmp,basile2016book}. The ME model rigorously explains the anomalous transport properties showing the slow decay in the current correlation $\beta=1/2$ \cite{basile2006prl,basile2009cmp}, super-diffusion in the energy diffusion \cite{jara2009,basile-olla-spohn-2010,jara2015}, and nonequilibrium steady current under finite thermal gradient \cite{lepri-monasterio-2009,basile2016book}. Herein, we extend the technique developed in \cite{basile2006prl,basile2009cmp} to the long-range interacting case with and without onsite pining potential. In particular, we focus on the current correlation. We present a brief summary for the one-dimensional systems in table \ref{tone}.
This table includes three important results. The first result is that the exponent $\beta$ in the current correlation {\it continuously} changes as a function of the index $\d$ in the long-range potential. The second result is that there is a regime where the current correlation {\it diverges} as increasing the system size with fixed time, and hence the exponent $\beta$ cannot be defined. The third result is that {\it even the systems with the onsite pining potential can show anomalous behavior}, i.e., $\beta <1$ owing to the long-range interaction. In addition, we extend the analysis to higher-dimensions, and we show that long-range interaction can induce anomalous exponent even in the {\it three-dimensional systems}. These exact findings show that the exponent in the anomalous transport can appear in various systems with long range interactions, even if the equilibrium thermodynamic properties such as the additivity and extensivity are satisfied.

This paper is organized as follows. In Sec. \ref{sect:model}, we introduce our model and explain some notations and definitions. In Sec. \ref{sect:Ct}, our main result about analytical solution of the energy current correlation is presented, and we derive the results listed in the Table\ref{tone}.
In Sec.\ref{sect:hd}, our analysis is extended to higher-dimension focusing on the momentum conserving systems. Finally, we summarize and discuss our results in Sec. \ref{sect:summary}.

\section{\label{sect:model}  Model}

\subsection{\label{subsect:hamiltonian}Long-range interacting Harmonic chain}
\begin{figure}[t]
\centering
\includegraphics[width=8.5cm]{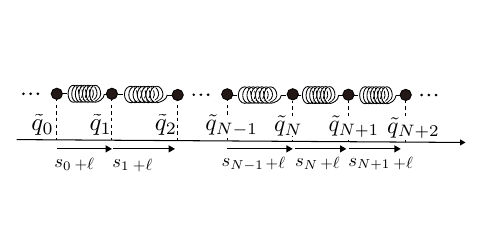}
\vspace*{-1.0cm}
\caption{\label{fig1} Schematic of the structure of spring system}
\end{figure}

We consider a classical one-dimensional system composed of $N$ particles. The position and momentum of the $x$th particle are denoted by $\tilde{q}_x$ and $p_{x}$, respectively. For the convenience of analysis, we consider the structure that is schematically shown in 
figure \ref{fig1}. That is, the infinite particles are arranged along the infinite line, and we focus on the dynamics of $N$ particles by imposing the boundary conditions appropriately. For momentum variables, we always impose the condition $p_x =p_{x+N}$. For the position variables, we impose different boundary conditions depending on whether the system has momentum conservation, as discussed below.

We employ the hybrid dynamics containing the deterministic dynamics from the Hamiltonian and stochastic exchange of momentum variables between the nearest neighbor sites, which is introduced in the subsequent subsection. The deterministic dynamics is induced by the following Hamiltonian that describes the long-range interacting harmonic chain:
\begin{align}
  H
 & =
\sum_{x} \f{p_{x}^{2}}{2} + {k_0\over 2} {(\tilde{q}_x -(x-1)\ell)^2 } +\sum_{x} \sum_{r=1}^{N/2}   V_{x,r}   \,  ,  \label{eq:H} \\
  V_{x,r}   & =   \f{1  }{ \tN r^{\d}} {(\tilde{q}_{x+r} - \tilde{q}_x - r \ell )^2 \over 2} \, , ~~~( \tN = \sum_{r=1}^{N/2}1/r^{\d} ) \, ,  \label{eq:H1}
\end{align}
where the index $\delta$ controls the range of the harmonic interactions. When $\d=\infty$, the interaction is reduced to the nearest neighbor interaction, while $\d\to 0$ implies global coupling. The factor $\tN$ is introduced to ensure extensivity on the equilibrium thermodynamic variables such as the free energy and entropy for $\d < 1$, whereas when $\d > 1$, the extensivity is satisfied even without this term. The quantity $\ell$ is the natural length of the springs.

\subsubsection{\label{sect:mcc}  Momentum-conserving case}
First, we explain the structure of the system without the onsite potential, i.e., $k_0=0$, called the {\it momentum-conserving case}. In this case, the spatially translational invariance is satisfied; hence, the total momentum is conserved. We define the stretch variable as
\begin{align}
s_x := \tilde{q}_{x+1} - \tilde{q}_x -\ell \, .  
\end{align}
Because we impose the boundary condition on the momentum $p_{x+N}\equiv p_{x}$, we immediately find that the summation of the stretch variables $\sum_{x'=x}^{N+x-1} s_{x'} = \tilde{q}_{N+x} - \tilde{q}_{x} - N \ell$ is conserved. Namely, for a given initial state, the length of the $N$ particles is conserved. For the momentum conserving systems, a specific value of $\tilde{q}_x$ does not matter because the global shift gives arbitrary values without changing the dynamics; hence, we impose the boundary condition for stretch variables as $s_{x+N}\equiv s_x$, and not for position variables. This boundary condition can be achieved once we set an initial configuration correctly. For each initial state, we define an average stretch as $\bar{s}=\sum_{x=1}^N s_x /N$. In addition, we introduce a displacement variable
\begin{align}
  q_x &:= \tilde{q}_x - (x-1) (\bar{s} + \ell ) \, .
\end{align}
Then, we find that $q_{N+x}=q_x$ automatically holds from the conservation of the length of $N$ particles. In this paper, for the momentum conserving case, we consider the initial states satisfying $\bar{s}=0$. In such initial states, the potential $V_{x,r}$ is rewritten as $V_{x,r}   =   (1/( \tN r^{\d} )) (q_{x+r}  - q_x  )^2/2 $.

In the momentum conserving case, there are three conserved quantities, total stretch, momentum, and energy. For the calculation of current correlation, we take a microcanonical average over the phase space with zero total stretch, zero total momentum, and a finite fixed total energy.

\subsubsection{Momentum-nonconserving case}
We next explain the structure of the system with the onsite potential, $k_0 \neq 0$, which we call the {\it momentum-nonconserving case}. In this case, there is a mechanical equilibrium position for each particle, i.e., $(x-1)\ell$ for the $x$th particle. In this system, a specific value in the position does matter; hence, we impose the boundary condition $\tilde{q}_{x+N}\equiv \tilde{q}_{x} + N \ell $ in addition to $p_{x+N}\equiv p_{x}$. It is convenient to introduce a displacement variable
\begin{align}
  q_x & := \tilde{q}_x - (x-1) \ell \, .
\end{align}
Then we find that $q_{x+N}= q_x$ is also satisfied. The potential $V_{x,r}$ is also rewritten as $V_{x,r}   =   (1/( \tN r^{\d} )) (q_{x+r}  - q_x  )^2/2 $.

In momentum-nonconserving systems, only the total energy is an important conserved quantity relevant to the detailed calculation of the current correlation. To calculate the current correlation, we take a microcanonical average over the phase space with a finite fixed energy. 

\subsubsection{Dispersion relation}
We consider the dispersion relation by which the sound velocity is defined. The dispersion relation and sound velocity are fundamental properties to characterize the macroscopic dynamics. A recent work \cite{tamaki2017} pointed out that the sound velocity can be important, especially in an open system attached to two reservoirs of different temperatures. Although in this study we focus on the current fluctuation in the closed setup, we list the classification of the sound velocities for different classes of system. 

Note that for momentum conserving and nonconserving cases, the potential term is reduced to the same expression with appropriately defined displacement variables $q_x$. We define the Fourier transform as follows
\begin{align}
  q_x &= {1\over\sqrt{N}} \sum_k q_k e^{-i  k  x} \, , \\
  p_x &= {1\over\sqrt{N}} \sum_k p_k e^{-i  k  x} \, , 
\end{align}
where the wave number is $k=2\pi/N, 4\pi/N,\cdots, 2 \pi$. In this study, the same Fourier transform is applied for different variables. Through straightforward calculation for the potential term, the dispersion relation can be obtained as
\begin{eqnarray}
\omega_{k}
=
\left[ k_0  +
\f{1}{\tN}
\sum_{r=1}^{N/2}
\f{4\sin^{2}{(kr /2)}}{r^{\d}}
\right]^{1/2} \, .
\label{omega}
\end{eqnarray}
In the proximity of $k=0$, this has the following asymptotic behavior
\begin{eqnarray}
\omega_{k}^2\sim
\begin{cases}
{\rm const.}\, ,&(0<\d < 1)\\
k_0 + a_{\delta} (\ln k^{-1} )^{-1} \, , &(\d = 1)\\
k_0 + a_{\delta}' \, k^{\d-1} \, ,&(1<\d < 3)\\
k_0 + a_{\delta}'' \, k^2\ln k^{-1} \, ,&(\d =3)\\
k_0 + a_{\delta}''' \, k^2 \, ,&(\d>3)
\end{cases} \, , 
\end{eqnarray}
where $a_{\delta}$, $a_{\delta}'$, $a_{\delta}''$, and $a_{\delta}'''$ are constants dependent on $\delta$. The sound velocity is given by the slope of $\omega_{k}$ at $k=0$. In the momentum-conserving case $k_0=0$, we find that the sound wave does not exist for $0<\d<1$, whereas the sound velocity is infinite for $1<\d \le 3$, and it is finite for $\d>3$. In the momentum-nonconserving case, the sound wave does not exist for $0<\d<1$, whereas the sound velocity is infinite for $1<\d < 2$, and it is zero for $\d>2$.

\subsection{\label{subsect:dynamics}Momentum-exchange dynamics with long-range interaction}
The dynamics is hybrid dynamics consisting of the deterministic part from the Hamiltonian and the stochastic part described by the random exchange of momentums between the nearest neighbor sites. For both momentum-conserving and nonconserving cases, the microscopic dynamics for variables $q_x$ and $p_x$ are the same. The infinitesimal change in the variables from time $t$ to $t+dt$ are described as follows.
\begin{align}
  d q_x &= p_x dt \, , \label{dynr1}  \\
  d p_x &= \left[ -k_0 q_x + \tN^{-1} \sum_{r=1}^{N/2} r^{-\d} ( q_{x+r} + q_{x-r} -2 q_x ) \right]dt \, \nonumber  \\
  &+ d n_x (p_{x+1} - p_x) + d n_{x-1}(p_{x-1} - p_x ) \, , \label{dynr2}
\end{align}
where $\{d n_{x}\}_{x=1}^{N}$ are independent stochastic variables, which take the value $0$ or $1$ with the Poisson process satisfying the noise average $\langle d n_x \rangle_{\rm n} =\gamma dt$. The noises stochastically exchange momentums between the nearest neighbor sites. This hybrid dynamics conserves total energy. In addition, for the momentum conserving case, the dynamics still satisfies the conservation of total momentum. When $\d$ is infinite, the interaction between the particles contains only nearest neighbor harmonic interaction; hence, the dynamics reduces to the original ME model discussed in Refs. \cite{basile2006prl,basile2009cmp,basile2016book}.

The corresponding dynamics for the distribution function can be obtained easily. Here, we only show for the momentum nonconserving case, where the distribution function for the phase space $({\bm q },{\bm p}):=(q_1,\cdots,q_N, p_1,\cdots, p_N)$ is defined \cite{ft1}. Let $P({\bm q },{\bm p},t)$ be the probability distribution. Because the stochastic noise is generated according to the Poisson process, the time evolution is given as
\begin{align}
  {\partial \over \partial  t}  P({\bm q},{\bm p},t)  &= (-\bA  +\g\bS) P({\bm q},{\bm p},t) \, , \label{distdyn}
\end{align}
where the operator $-\bA$ denotes the deterministic dynamics given by Liouville's operator
\begin{eqnarray}
\bA 
&:=&
\sum_{x=1}^{N}\l(
\f{\p H}{\p p_{x}}
\f{\p}{\p q_{x}}
-
\f{\p H}{\p q_{x}}
\f{\p}{\p p_{x}}
\r). \label{aexp}
\end{eqnarray}
The operator $\g\bS$ is the part of the stochastic dynamics that acts as
\begin{eqnarray}
\bS  P({\bm q},{\bm p},t) \, 
&:=&
     \sum_{x=1}^{N} \left[ P({\bm q},{\bm p}^{x|x+1} ,t) - P({\bm q},{\bm p},t)  \right] \, ,  \label{bsexp}
\end{eqnarray}
where ${\bm p}^{x|x+1}$ is obtained by substituting $p_x,p_{x+1}$ with $p_{x+1},p_x$ in ${\bm p}$.

\subsection{\label{subsect:current}Energy current}
Energy current is defined by the continuity equation of the local energy. Therefore, we define the local energy as
\begin{eqnarray}
\e_{x}
=
\f{p_{x}^{2}}{2}
+
\f{1}{2\tN}
  \sum_{r=1}^{N/2}\left[
\frac{(q_{x+r}-q_{x})^{2}}{2r^{\d}}
+
\frac{(q_{x}-q_{x-r})^{2}}{2r^{\d}}
\right]\, . ~~
\end{eqnarray}
The evolution of the local energy is calculated according to (\ref{dynr1}) and (\ref{dynr2}). We need to be careful as the time evolution involves stochastic terms. In addition, we also note that the dynamics contains nonlocal interaction, which inevitably leads to nonlocal expression on the energy current. First, we consider the infinitesimal change in the local energy:
\begin{align}
  d \e_{x}
           &={1\over \tN} \sum_{r=1}^{N/2}{1\over 2 r^{\d}}\left[
             -  (q_x - q_{x+r})(p_{x+r} + p_x) dt  \right. \nonumber \\
           &\left.  ~~~~~~~~~~~~~~~~~+ (q_{x-r} - q_x) (p_{x-r} + p_x)  dt   \right] \nonumber \\
           &+dn_x \left( {p_{x+1}^2 \over 2} - {p_{x}^2 \over 2} \right)+dn_{x-1} \left( {p_{x-1}^2 \over 2} - {p_{x}^2 \over 2} \right) \, ,
\end{align}
where the expression containing the noise terms denote the exchange of kinetic energies caused by the exchange of momentums between the nearest neighbor sites.

\begin{figure}[t]
\centering
\includegraphics[width=8.5cm]{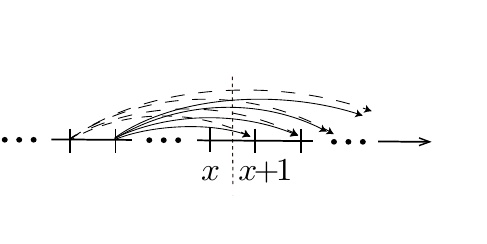}
\vspace*{-1.0cm}
\caption{\label{fig2} Interpretation of energy current $j^{\rm A}_{x}$. Each arrow indicates the direct transmission of energy from one site to another. The energy current is defined by counting all transmissions through the surface between the site $x$ and $x+1$.}
\end{figure}

Next, we compare the above expression with the continuity equation with respect to energy $d \epsilon_x = -d j_x + d j_{x-1}$. Note that the Hamiltonian satisfies translational invariance; hence, the current expressions should be constructed such that the expressions of $d j_x$ and $d j_{x-1}$ are identical to each other once we shift the site index. From this criterion, we can derive the following expression of energy current:
\begin{align}
dj_{x}
&:= (j^{\rm A}_{x}+\g j^{\rm S}_{x})dt+ d\j_{x} \, , \label{jx} \\
j^{\rm A}_{x} 
&:=
-\f{1}{\tN}
\sum_{x'=x+1}^{x+N/2}
\sum_{r=x'-x}^{N/2}
\f{q_{x'}-q_{x'-r}}{r^{\d}}
\f{p_{x'}+p_{x'-r}}{2}, \label{jax}\\
j^{\rm S}_{x}
&:=
-\f{p_{x+1}^{2}-p_{x}^{2}}{2},\\
d\j_{x}
&:=
-\f{p_{x+1}^{2}-p_{x}^{2}}{2}d\m_{x}.
\end{align}
Here, $d\m_{x}$ is the Martingale noise defined as $d\m_{x}:=d n_{x}-\gamma dt$ \cite{protter}. The currents $j^{\rm A}_x$ and $j^{\rm S}_x$ are the instantaneous currents from the deterministic dynamics and average stochastic noise, respectively. The third current $d\j$ is a current from the Martingale noise. Note that the expression of $j^{\rm A}_x$ is nonlocal, which is a direct consequence of long-range interactions. By  considering all contributions of the energy transmissions across the {\it surface} between the sites $x$ and $x+1$, the expression $j^{\rm A}_x$ is defined. Figure \ref{fig2} shows the schematic for the interpretation.

%

\section{\label{sect:Ct}ENERGY CURRENT CORRELATION}
\subsection{Definition of the current correlation}
We consider the energy current correlation that is directly related to thermal conductivity via the Green--Kubo formula. From the current expressions, we note $j_{\rm tot}:=\sum_{x=1}^{N} j_{x}^{\rm A} + j_{x}^{\rm S}=j_{\rm tot}^{\rm A}$, where $j_{\rm tot}^{\rm A}:=\sum_{x=1}^{N} j_{x}^{\rm A}$. Then, we consider the following current correlation \cite{basile2006prl,basile2009cmp,tamaki2017,saito-sasada-2017}:
\begin{align}
C_{N}(t)
  &:= \f{1}{N} \Bigl\langle  j_{\rm tot}(t)j_{\rm tot} \Bigr\rangle_{\rm mc:n} \,  \nonumber  \\
  &= \Bigl\langle  j_{\rm tot}^{\rm A}(t)    j_{0}'    \Bigr\rangle_{\rm mc:n} .
\label{eq:def-Ct}
\end{align}
where the symbol $\la ...  \ra_{\rm mc:n}$ denotes the microcanonical average (mc) as well as the noise average (n). The variable $ j_{0}'$ is defined by noting that the total current $j_{\rm tot}^{\rm A}$ is simplified from the expression of local current (\ref{jax}) and is rewritten with the new variable $ j_{x}'$ as
\begin{align}
  j_{\rm tot}^{\rm A}&=\f{-1}{\tN} \sum_{x} \sum_{r=1}^{N/2}  \f{q_{x}-q_{x-r}}{r^{\d -1}} \f{p_{x}+p_{x-r}}{2}                       =\sum_{x} j_{x}' \, , \\
  j_{x}' &:= \f{-1}{\tN} \sum_{r=1}^{N/2}  \f{q_{x}-q_{x-r}}{r^{\d -1}} \f{p_{x}+p_{x-r}}{2} \, . 
\end{align}
Using the translational invariance in the system, we selected one site in Eq.(\ref{eq:def-Ct}).

The current correlation is directly related to the thermal conductivity via the Green--Kubo formula, if the time integral is finite \cite{kth}. In anomalous heat transport, the combination of power law decay in the current correlation and time integration up to the cut-off time $N/c$, where $c$ is the sound velocity, is thought to explain the system size dependence of the diverging thermal conductivity.

Note that the current expression (\ref{jx}) also has the Martingale part. However, it is known that its contribution to the thermal conductivity is constant, and the correlations between the Martingale part and $j^{\rm A}, j^{\rm S}$ vanish \cite{basile2006prl,basile2009cmp,basile2016book}. Hence, we do not involve the Martingale current in the definition of the current correlation.

\subsection{ Laplace transform of $C_N(t)$}
We outline the calculation below. First, we consider the Laplace transform:
\begin{align}
\hC_{N}(\lam)
& :=
\int_{0}^{\infty}dt\,
e^{-\lam t}C_{N}(t) \, \nonumber \\
&= \int_{0}^{\infty} dt
e^{-\lam t}
\langle 
                                      [e^{(\bA+\g\bS)  t} j_{\rm tot}^{\rm A} ] \,  j_{0}'
                            \rangle_{\rm mc} \,  , \nonumber  \\ 
&=
                                                                    \la[(\lam\!-\!\bA\!-\!\g\bS)^{-1} j_{\rm tot}^{\rm A}                                                                    ] \,  j_{0}'  \rangle_{\rm mc} \, , \label{clap}
\end{align}
where, from the second line, the expression is the form by taking the noise average. The dynamics for the variables is given by the operator $\bA$ and $\bS$; the dynamics is conjugate to the distribution function (\ref{distdyn}). 

For further calculation (\ref{clap}), we consider the equation
\begin{align}
(\lam-\bA-\g\bS)u(\lam)
= j_{\rm tot}^{\rm A} \, .
\label{eq:resolvent}
\end{align}
Note that using the quantity $u(\lam)$, the Laplace transform of the current correlation can be written as $\hC_{N}(\lam)= \la u(\lam)  j_{0}' \ra_{\rm mc}$.
To obtain the explicit expression of $u(\lam)$, we impose the following form
\begin{eqnarray}
u(\lam)
=
\sum_{x,x'}g_{x-x'}  q_{x}p_{x'} \, , 
\end{eqnarray}
where we also assume the relation $g_{-x}=-g_{x}$ and $g_{x+N}=g_{x}$. Substituting this expression into Eq.(\ref{eq:resolvent}) and comparing the coefficients of the term $q_x p_{x'}$, the relation is satisfied:
\begin{align}
  \left( \lambda - \gamma \Delta_{x' } \right) g_{x-x'}  &={-1\over 2 \tN r^{\d -1}}\left( \bar{\delta}_{x-x' , r}- \bar{\delta}_{x-x' , -r}    \right) \, , 
\end{align}
where $\bar{\delta}_{x,a}$ is the Kronecker's delta function, i.e., $\bar{\delta}_{x,a}=1$ for $x=a$, and $\bar{\delta}_{x,a}=0$ otherwise. The symbol $\Delta_{x'}$ is the discrete Laplacian that acts as $\Delta_{x'} f_{x'} := f_{x'+1}+ f_{x'-1}-2f_{x'}$. 
Through the Fourier transform for both sides in the equation, the explicit form of the function $g_{x-x'}$ is easily obtained as
\begin{align}
  g_{x-x'} &= {1\over \sqrt{N} } \sum_k g_k  e^{-i k (x - x' )} \, , \\
  g_k &=  {-i \over \sqrt{N} }  { \Phi_k \over  \lambda + \gamma \left[ 2 \sin (k/2) \right]^2 } \, , 
\end{align}
where
\begin{align}
  \Phi_{k}  &:= \f{1}{\tN}\sum_{r=1}^{N/2}\f{\sin{(k r)}}{r^{\d-1}}.
\label{eq:Phi}             
\end{align}
The function $\Phi_{k}$ is related to the Fourier representation of the total instantaneous current as $j_{\rm tot}^{\rm A}=-i\sum_k \Phi_{k} q_{-k} p_k$.

Finally, we consider taking the average over the microcanonical average, where we proceed computation based on the ensemble equivalence between microcanonical and canonical distribution. Note the expression
\begin{align}
  \hC_{N} (\lam) &=
                   \sum_{r, x,x'}
                   {-1 \over 2 \tN r^{\delta -1}} g_{x -x'} \langle q_x p_{x'} (q_0 - q_{-r})(p_0 + p_{-r})\rangle_{\rm mc}  \nonumber \\
                 &=\sum_{r,x ,x'}{-1\over 2 \tN r^{\delta -1}} g_{x - x'}   \nonumber  \\
                 &~~~~~~~~~~~ \times   \langle q_x  (q_0 - q_{-r}) \rangle_{\rm mc} \langle p_{x'} (p_0 + p_{-r})\rangle_{\rm mc} \nonumber \\
          &={-\langle p_0^2\rangle_{\rm mc} \over 2 \tN} \sum_{r}{1\over r^{\delta -1}} \sum_{x} g_{x }  h_r(x)  \, , \\
h_r (x) & :=\langle q_x (q_r - q_{-r})\rangle_{\rm mc} \, , 
\end{align}
where to obtain the function $h_r (x)$, we used the translational invariance to shift the site index in the correlation. For the expression of the function $h_r (x)$, we further use the ensemble equivalence between the microcanonical and canonical ensemble (the detailed calculation is provided in the Appendix), and we have the following expression for the Fourier transform
\begin{align}
  h_r (k) &= {  4i k_{\rm B} T \over \sqrt{N}}  { \sin (k r) \over \omega_k^2 } \, .  
\end{align}
The Fourier transform of the resultant expression of $\la u(\lam) j_0' \ra_{\rm mc}$ is as follows:
\begin{align}
  \hC_{N}(\lam)&= {2 (k_{\rm B} T)^{2} \over N} \sum_k  { 1 \over  \lambda + \gamma \left[ 2 \sin (k/2) \right]^2 }
                 {\Phi_k^2 \over \omega_k^2}  \, \nonumber \\
  &\sim {2 (k_{\rm B} T )^{2}\over \pi}\int_{N^{-1}}^{\pi} dk  { 1 \over  \lambda + \gamma \left[ 2 \sin (k/2) \right]^2 }
                 {\Phi_k^2 \over \omega_k^2}  \, , 
                   \label{cjj}
\end{align}
where we take the continuous expression in terms of the wave number in the last line. We reduced the interval of integration to $\left[ N^{-1} , \pi \right]$ using the symmetry with respect to $k= \pi$. The constant $k_{\rm B}$ is the Boltzmann constant.

\subsection{\label{asym}Asymptotic behavior of current correlation}
Now, we can analyze the asymptotic behavior of the current correlation.  The correlation function in the time domain is obtained with the inverse Laplace transform
\begin{align}
  C_N (t) &= {1\over 2 \pi i} \int_{c-i\infty}^{c+ i\infty} d \lambda  \,   \hC_{N}(\lam) e^{\lambda t} \,  \nonumber \\
  &={2 (k_{\rm B} T)^2 \over \pi } \int_{N^{-1}}^{\pi} dk \, {\Phi_k^2 \over \omega_k^2} e^{-\gamma \left[ 2 \sin (k/2)\right]^2 t } \, . 
\end{align}
Here, we have selected the pole $- \gamma \left[ 2 \sin (k/2) \right]^2 $ in the $\lambda$-plane. From this, one can recognize that the asymptotic behavior in time is obtained from the small wave number regime. The asymptotic behavior in the time domain considering the behavior of small wave numbers is discussed below.

\subsubsection{Momentum-conserving case}
In the short-range interacting case, i.e., in the ME model, it is exactly shown that the exponent of the time-decay in the current correlation function is $1/2~(<1)$, which implies anomalous transport. Now we consider the long-range interacting case satisfying total momentum conservation \cite{basile2006prl}. 

For small $k$, the function $\Phi_{k}^{2}/\omega_{k}^{2}$ behaves as
\begin{eqnarray}
\f {\Phi_{k}^{2}}{\omega_{k}^{2}}
\sim
  \begin{cases}
    k^{-2} \, , &(\d < 1) \\
    k^{-2} (\ln k^{-1})^{-1}\, , &(\d = 1) \\
  k^{-(3-\d)}\, , &(1<\d<3)\\
  \ln k^{-1} \, , &(\d=3)\\
{\rm const.}\, ,&(\d < 3)
\end{cases}.
\end{eqnarray}
Hence, for $\d\le 2$, the integral in Eq. (\ref{cjj}) exhibits infrared divergence in the limit of $N\to\infty$. The asymptotic behavior is expressed as
\begin{eqnarray}
\lim_{N\to\infty}C_{N}(t)
\sim
  \begin{cases}
    N \, , &(\d <1) \, \\
    N^{2-\d}\, ,&(1< \d<2)\\
    \ln N \, , &(\d =2) \\
C(t) \, ,&(\d>2)
\end{cases} \, , 
\label{eq:limN-C}
\end{eqnarray}
where $C(t)$ is the saturated continuous function in the thermodynamics limit. The current correlation is ill-defined in the thermodynamic limit for $\d<2$. Note that the appropriately scaled current correlation, i.e., $C_{N}/N$ for $0<\d<1$ and $C_{N}/N^{2-\d}$ for $1<\d<2$, is independent of $t$ for large $t$. From the structure of the inverse Laplace transform, we can find the asymptotic behavior of the function $C(t)$ given for $\d>2$:  
\begin{eqnarray}
\lim_{t\to\infty}
C(t)
\sim
\begin{cases}
  t^{-(\d-2)/2}\, ,&(2<\d<3)\\
  t^{-1/2} \ln t \, , & (\d = 3) \\
t^{-1/2}\, ,& (\d > 3)
\end{cases}.
\label{eq:limt-C}
\end{eqnarray}
This result leads to the classification of the exponent $\beta$ in Eq.(\ref{ctintro}) and the results listed in table \ref{tone}. A crucial observation here is that the exponent of the asymptotic time-decay {\it continuously} changes as a function of the index of long-range potential $\d$. Another crucial observation is that there is an ill-defined regime ($\d < 2$) where the current correlation diverges. The exponent $1/2$ for $\d > 3$ implies that this regime is regarded as a short-range interaction in the context of heat transfer. 

\subsubsection{Momentum-nonconserving case}
Next, we consider the momentum-nonconserving case, i.e., $k_0\neq 0$. We note that in the short-range interacting case, the onsite potential induces normal thermal conduction. It is already known that the exponent in time-decay in the current correlation is $3/2~ (>1)$ \cite{basile2006prl}. Here, we consider the effect of long-range potential based on the exact expression. We perform an analysis similar to the one in previous subsection. For the term $(\Phi_{k}^{2}/ \omega_{k}^{2} )$, the main contribution for the small wave number regime is from $\Phi_{k}^{2}$, because the dispersion relation is constant for the regime. Then, we have 
\begin{eqnarray}
\f {\Phi_{k}^{2}}{\omega_{k}^{2}}
&\sim
                                    \begin{cases}
                                      k^{-2} \, , & (\d < 1) \\
                    (k \ln k^{-1})^{-2} \, , & (\d =1) \\                  
  k^{2(\d -2)} \, ,&(1< \d<3)\\
   (k \ln k^{-1} )^2  \, ,& (\d= 3)\\
 k^2  \, , &(\d > 3)
\end{cases}.
             \label{eq:limt-Cnc}
\end{eqnarray}
From this expression, we can discuss the thermodynamic limit by considering the system-size dependence:  
\begin{eqnarray}
       \lim_{N\to\infty}C_{N}(t)
&\sim
                                   \begin{cases}
                                     N \, , & (\d < 1) \\
                                     N^{-(2\d -3)}\, , &(1<\d<3/2)\\
                                     \ln N \, , &(\d =3/2) \\
C(t) \, ,&(\d >3/2)
\end{cases}.
\end{eqnarray}
From this, we find that there is a regime that the current correlation shows infrared divergence ($\d < 3/2$). For $\d >3/2$, we have the well-defined continuous function $C(t)$. The asymptotic behavior in the time-domain of the function is estimated as 
\begin{eqnarray}
\lim_{t\to\infty}
C(t)
\sim
\begin{cases}
t^{-(2\d-3)/2}\, ,&(3/2<\d<3)\\
t^{-3/2} ( \ln t )^2\, , &(\d = 3) \\
t^{-3/2}\, .&(\d > 3)
\end{cases}.
\label{eq:limt-C:onsite}
\end{eqnarray}
This leads to the classification of the exponent $\beta$ in the time-decay listed in table \ref{tone}. A critical observation here is that the exponent can be less than $1$ for $3/2 < \d < 5/2$, which indicates anomalous behavior in heat conduction. This is physically important because the momentum-nonconserving systems have been thought to show normal heat conduction. This anomalous behavior originates from the long-range interaction. Hence, one can say that {\it long-range interactions induce the anomaly}. Again, we observe that the exponent continuously changes as a function of the index of the long-range interaction $\d$. For $\d > 3$, we have the exponent $3/2$, which is the same as in the short-range interacting case. Considering the exponent, the regime $\d >3$ can be regarded as a short-range interaction. On the other hand, in the context of the normal heat conduction, the index $\d = 5/2$ is critical since the exponent $\beta$ is larger than $1$ for $\d  > 5/2$. 

\section{\label{sect:hd}Extension to high-dimensions}
We herein consider the effect of higher dimensions. Having computed the one-dimensional systems, it is now straightforward to extend the calculation to higher dimensions. It is known that in short-range interacting momentum-conserving systems, the dimensionality significantly affects on the long-time tail of current correlation, which in general leads to the normal thermal conductivity in the three dimension. This has been checked numerically for the FPU lattices \cite{saitodhar}, and the exact analysis for the three-dimensional extension of the ME model showing the convergence of the Green-Kubo integral in \cite{basile2006prl}. We also remark that the transport behavior in the two-dimensional systems seem to show several varieties depending on the dynamics \cite{saitodhar, basile2006prl,pre2010}. Given these observations for short-range interacting systems, we here examine what happens in the long-range interacting systems within the present model. High-dimensionality tends to enhance a fast relaxation while long-range interaction tends to induce slow relaxation. To focus on such competition, we here consider only the momentum conserving systems. As shown below, even the three-dimensional systems can be anomalous in the exponent of the time-decay due to the long-range interaction. 

\subsection{Dynamics}
Let us consider the $d$-dimensional hiper-cubic lattice with the size $N^d$. We assign $d$-dimensional vector for displacement variables and momentum variables. 
The Hamiltonian is described as
\begin{align}
  H &= \sum_{{\bm x} \in \mathbb{Z}_{N}^d} {|{\bm p}_{\bm x}|^2 \over 2} + \sum_{{\bm r} \in {\bm I}_N^d  }
      V( |\tilde{\bm q}_{{\bm x}+ {\bm r}} - \tilde{\bm q}_{\bm x} -\ell {\bm r}      | ) \, ,
\end{align}
where ${\bm x}$ labels a site $\in \mathbb{Z}_{N}^d$ where $\mathbb{Z}_{N}^d =\mathbb{Z}^d/ N\mathbb{Z}^d$. The variables ${\bm p}_{\bm x}$ and $\tilde{\bm q}_{\bm x}$ are respectively the momentum and position variable at the site ${\bm x}$. The term $V$ is a long-range spring potential between the position ${\bm x}$ and ${\bm x} + {\bm r}$ where ${\bm r}$ is a relative vector taken from the set
\begin{align}
  {\bm I}_N^d &= \left\{ {\bm r} =\sum_{i=1}^d \nu_i {\bm e}_i \,| \, \nu_i=0,1,\cdots , N/2 ~ \backslash \{ {\bm 0}\}\right\} \, ,
\end{align}
with the unit vector ${\bm e}_i~(i=1,\cdots, d)$. The detailed expression of the long-range potential term is written as
\begin{align}
  V( |\tilde{\bm q}_{{\bm x}+ {\bm r}} - \tilde{\bm q}_{\bm x} -\ell {\bm r} | ) &={1\over 2  \tilde{N} } { | \tilde{\bm q}_{{\bm x} + {\bm r} } - \tilde{\bm q}_{\bm x} -\ell {\bm r}|^2 \over  |{\bm r}|^{\delta}} \, , \\
  \tilde{N} &= {1\over d} \sum_{{\bm r} \in {\bm I}_N^d } {1\over |{\bm r} |^{\delta}} \, . 
\end{align}
Now we take the same procedure as in Sec.{\ref{sect:mcc}}. First we impose ${\bm p}_{{\bm x}+N{\bm e}_i}={\bm p}_{{\bm x}}$ for $i=1,\cdots,d$. Next, introducing the stretch variable ${\bm s}_{{\bm x}}^{(i)} = \tilde{\bm q}_{{\bm x}+{\bm e}_i} - \tilde{\bm q}_{{\bm x}}  - \ell {\bm e}_i$, we impose the boundary condition ${\bm s}_{{\bm x}}^{(i)}= {\bm s}_{{\bm x}+N{\bm e}_i}^{(i)}$. As in the one-dimensional case, we consider the microcanonical ensemble with the configuration space $\sum_{j=1}^{N} {\bm s}_{{\bm x}+j {\bm e}_i }^{(i)} =0$. In addition, we define the new displacement variable:
\begin{align}
  {\bm q}_{\bm x} &:= \tilde{\bm q}_{\bm x} - ({\bm x} - {\bm 1})\ell  \, ,
\end{align}
where ${\bm 1}=(1,\cdots , 1)$. This variable satisfies ${\bm q}_{\bm x} = {\bm q}_{{\bm x} + N{\bm e}_i}$. 

Introducing the exchange noise between the nearest neighbor sites, $n_{{\bm x}, {\bm x}+{\bm e}_i}$, the dynamics for infinitesimal time step is given as follows
\begin{align}
  d q_{{\bm x},i}  &= p_{{\bm x},i} dt \, , \\
  d p_{{\bm x},i} &=  \tilde{N }^{-1} \sum_{{\bm r} \in {\bm I}_N^d  } |{\bm r}|^{-\delta }  ( q_{{\bm x} +{\bm r} ,i} + q_{{\bm x} -{\bm r} ,i} - 2 q_{{\bm x} ,i} )  dt
                    \, \nonumber \\
                  & +\sum_{j=1}^d \Bigl[ dn_{{\bm x}, {\bm x}+{\bm e}_{j}  } ( p_{{\bm x}+{\bm e}_{j}, i } - p_{{\bm x}, i } ) \nonumber \\
  &~~~~~~~~~~~~~~~+
                    dn_{{\bm x}, {\bm x}-{\bm e}_{j}  } ( p_{{\bm x}-{\bm e}_{j}, i } - p_{{\bm x}, i } )
                    \Bigr] \, , 
\end{align}
where $q_{{\bm x},i}$ and $p_{{\bm x},i}$ are respectively the $i$th component of the vectors ${\bm q}_{{\bm x}}$ and ${\bm p}_{{\bm x}}$. The exchange noises obey the Poisson statistics, i.e., $\langle dn_{{\bm x}, {\bm x}+{\bm e}_j} \rangle_{\rm n} = \gamma dt$.

\subsection{Current correlations}
We set the local energy as
\begin{align}
  \epsilon_{\bm x} &:= {|{\bm p}_{\bm x}|^2 \over 2} + {1\over 2 \tilde{N}}\sum_{{\bm r} \in {\bm I}_N^d  } \Bigl[ 
                     { | {\bm q}_{{\bm x} + {\bm r} } - {\bm q}_{\bm x} |^2 \over  2 |{\bm r}|^{\delta}}
                     +  { |  {\bm q}_{\bm x} - {\bm q}_{{\bm x} - {\bm r} } |^2 \over  2 |{\bm r}|^{\delta}}
                     \Bigr]
                     \, .
\end{align}
Then, through the continuity equation with respect to energy, one can identify the energy current expression, and we eventually arrive at the following expression for the current correlation
\begin{align}
  C_{N}^{(i,i)} (t) &= {1\over N^d} \langle j_{{\rm tot},i}^{\rm A} (t) j_{{\rm tot},i}^{\rm A}  \rangle_{\rm mc:n} \nonumber \\
  &=  \langle j_{{\rm tot},i}^{\rm A} (t) j_{{\bm 0},i}'  \rangle_{\rm mc:n}  \, , \\
  j_{{\rm tot},i}^{\rm A} &=\sum_{\bm x} j_{{\bm x},i}' \, , \\
  j_{{\bm x},i}' &= {1\over \tilde{N}}
                   \sum_{{\bm r} \in {\bm I}_N^d  }{  r_i (  q_{{\bm x} -{\bm r} ,i} - q_{{\bm x} ,i}   )\over |{\bm r}|^{\delta } }
                   {  (  p_{{\bm x} ,i} + p_{{\bm x} -{\bm r} ,i} )\over 2 }
                   \, ,
\end{align}
where the superscript $(i,i)$ means that we consider the correlation between currents in the $i$th direction only. The computation is completely in parallel to the one-dimensional case. The resultant expression reads
\begin{align}
  C_{N}^{(i,i)} (t) &={2d (k_B T)^2 \over N^d} \sum_{{\bm k}} { (\Phi_{\bm k}^{(i)}  )^2 \over \omega_{\bm k} ^2}  e^{-t \gamma \sum_{j=1}^d   \left[ 2 \sin ( {\bm k}\cdot {\bm e}_j /2) \right]^2}  \, , 
\end{align}
where ${\bm k}$ is the $d$-dimensional wave length vector, i.e. the $i$th component is $k_i = {2 \pi /N},{4 \pi /N},\cdots , 2 \pi $. The term $\omega_{\bm k}$ is a dispersion relation. The detailed expression of $\omega_{\bm k}$ and $\Phi_{\bm k}^{(i)}$ are given as follows.
\begin{align}
  \omega_{\bm k}^2 &= {1\over \tilde{N}}  \sum_{{\bm r} \in {\bm I}_N^d  }   { \left[ 2 \sin ({\bm k}\cdot {\bm r} /2 ) \right]^2 \over |{\bm r}|^{\delta}}   \, , \\
  \Phi_{\bm k}^{(i)} &= {1 \over \tilde{N} }\sum_{{\bm r} \in {\bm I}_N^d  }  {r_i \over |{\bm r}|^{\delta}    } \sin ({\bm k} \cdot {\bm r} ) \, . 
\end{align}
\subsection{Asymptotic behavior}
In order to consider the long-time tail, we note the expressions at small wave numbers for $ (\Phi_{\bm k}^{(i)}  )^2 / \omega_{\bm k}^2$:
\begin{eqnarray}
  {(\Phi_{\bm k}^{(i)}  )^2 \over  \omega_{\bm k}^2 } &\sim &
                                                              \left\{
              \begin{array}{ll}
                k^{-2}, &(\delta < d)  \\
                k^{-2}/\ln k^{-1}  , & (\delta =d) \\
                k^{-(d +2 -\delta)} , &(d < \delta < d+2 ) \\
                \ln k^{-1} , &(\delta = d+2) \\
                {\rm const.} , & (\delta > d +2) 
                \end{array}
                                 \right. \, ,
\end{eqnarray}
We should also note 
\begin{eqnarray}
  \tilde{N} &\sim & \left\{
              \begin{array}{ll}
                k^{-(d-\delta )}, &(\delta < d)  \\
                \ln k^{-1}  , & (\delta =d) \\
                {\rm const.} , & (\delta > d)
                \end{array}
                                 \right. \, .
\end{eqnarray}
Then, through the similar computation to the one-dimensional case, we can obtain the following dimension dependence:
\begin{eqnarray}
d=1: \lim_{N\to \infty} C_{N}^{(i,i)} (t) &=& \left\{
                                                \begin{array}{ll}
                                                  \infty \, , & ( \delta \le 2) \\
                                                   t^{-(\delta -2)/2} \, , & (2 < \delta < 3) \\
                                                  t^{-1/2} \ln t \, , & (\delta = 3) \\
                                                    t^{-1/2} \, , & (\delta > 3)
                                                \end{array}
                                                                \right. , \nonumber  \\
  && \\
  d=2: \lim_{N\to \infty} C_{N}^{(i,i)} (t) &=& \left\{
                                                \begin{array}{ll}
                                                  \infty \, , & ( \delta \le 2) \\
                                                   t^{-(\delta -2)/2}\,  , & (2 < \delta < 4) \\

t^{-1}\ln t \,  , & (\delta = 4) \\
                                                   t^{-1}\,  , & (\delta > 4)
                                                \end{array}
                                                              \right. , \nonumber \\
                                           && \\
  d \ge 3: \lim_{N\to \infty} C_{N}^{(i,i)} (t) &=& \left\{
                                                \begin{array}{ll}
                                                  t^{-(d-2)/2} \, , & ( \delta \le d) \\
                                                   t^{-(\delta -2)/2} \, , & (d < \delta < d+2) \\
                                                  
t^{-d/2} \ln t \, , & (\delta = d+2) \\

 t^{-d/2} \, , & (\delta > d+2)
                                                \end{array}
                                                              \right. . \nonumber \\  
\end{eqnarray}
Now we see that even in high-dimensional systems, the current correlation can be anomalous in the sense of (\ref{ctintro}). Remarkably, even the three-dimensional systems can be anomalous due to the long-range interaction for the parameter region $\delta <4$.

\section{\label{sect:summary}SUMMARY and DISCUSSION}
In this study, we consider the effect of long-range interaction in the energy current correlation. The current correlation is a key component in the Kubo formula leading to thermal conductivity. To obtain clear-cut results, we introduce the exactly soluble model, which reduces to the momentum exchange model in the short-range interaction limit, and we derive the exponent $\beta$ in Eq.(\ref{ctintro}) exactly.

We compare the momentum conserving case with nonconserving case, because it is known that the momentum-conserving case with a short range interacting case shows anomalous transport with the exponent $\beta=1/2 ~ (<1)$, whereas the momentum-nonconserving case does not show an anomaly, i.e., $\beta=3/2 ~ (>1)$. In terms of the index of long-range interaction $\d$, the results of the exponents are summarized in table \ref{tone}. We have three main results. First, the exponent $\beta$ continuously changes as a function of the index of the long-range potential $\d$. Second, there is a regime where the current correlation function is ill-defined. Finally, the most remarkable finding is that even momentum-nonconserving case can exhibit anomalous transport for a certain range of $\d$.
We note that recent paper \cite{xiong} carefully discusses the effect on the scaling factor $\tilde{N}$. In the present model, the essential results do not depend on the existence (or absence) of the scaling factor. These observations might be suggestive for realistic experiments with physical objects that involve long-range terms in potentials. 

We also extend the one-dimensional analysis to the high-dimension focusing on the momentum conserving case. Then we exactly show that the long-range interaction can induce the anomalous exponent even in the three-dimensional systems. This is another important message in this paper. 

In this paper, we employ the toy model with stochastic noises to solve the current fluctuation exactly, which is definitely useful to understand anomalous behavior in the transport. In the nonlinear dynamics, however, there are many unsolved problems that originate from the nonlinearity in the dynamics. For short range interacting cases, many types of interactions are investigated to study the transport properties. Recently, fluctuating hydrodynamics has a central role to analyze transport phenomena in short-range interacting systems \cite{beijeren2012,spohn2014,das2014}. On the other hand, it is not yet clear how to connect the fluctuating hydrodynamics to the nonlinear long-range interacting systems. It is a crucial future problem to find the connection to understand the underlying mechanism of the transport.

\begin{acknowledgments}
We are grateful to Makiko Sasada and Hayate Suda for fruitful discussions. The present work was supported by JSPS Grants-in-Aid for Scientific Research (JP16H02211 and JP17K05587).

\end{acknowledgments}

\appendix

\section{\label{app:Ct} Calculation of  $h_r (x)$}
We consider the microcanonical average for the function $h_r (x)$:
\begin{align}
h_r (x) & :=\langle q_x (q_r - q_{-r})\rangle_{\rm mc} \, . 
\end{align}
We consider this function for momentum-conserving and nonconserving cases separately.

First, we consider the momentum-nonconserving case. Note that the function $h_r (x)$ is the correlation on the local observables.
Calculating the local observable in terms of the microcanonical ensemble is equivalent to calculating the expectation value of the local observable in terms of the locally reduced distribution function from the microcanonical ensemble. Now, we impose the ensemble equivalence between the microcanonical ensemble with a fixed energy and the canonical ensemble with the corresponding temperature $T$:
\begin{align}
  \langle q_x (q_r - q_{-r})\rangle_{\rm mc} &\sim   \int d\Gamma  q_x (q_r - q_{-r})  \rho_{\rm can} ({\bm q} , {\bm p}) \, \\
   \rho_{\rm can} ({\bm q} , {\bm p}) &= \exp \left( - H/(k_{\rm B} T)  \right) /Z_{T}  \, ,
\end{align}
where $\int d\Gamma ...$ is the phase space average, i.e., $\int dq_1 dq_2 \cdots dp_1 dp_2 \cdots $. The function $Z_{T}$ is the partition function. The boundary condition for this case is $q_{x+N}=q_x$. We obtain the expression as
\begin{align}
  h_r (x) &= {1\over \sqrt{N}}\sum_k h_r (k) e^{-i k x} \, , \\
  h_r (k) &= {  4i k_{\rm B} T \over \sqrt{N}}  { \sin (k r) \over \omega_k^2 } \, .  \label{hrk1}
\end{align}

Next, we consider the momentum-conserving case. In this case, we use the phase space $({\bm s} , {\bm p})$ instead of $({\bm q},{\bm p})$. Hence, we note the following argument on the function $h_r (x)$. Using the discrete Laplacian, we have the following expression: 
\begin{align}
  \Delta h_r (x) & = \langle (s_x - s_{x-1})(q_r - q_{-r} ) \rangle_{\rm mc} \, \nonumber \\
                 &= \langle (s_x - s_{x-1})(s_{r-1} + s_{r-2} + \cdots + s_{-r} ) \rangle_{\rm mc} \, \nonumber \\
                 &=  \langle s_x (s_{r-1} + s_{r-2} + \cdots + s_{-r} ) \rangle_{\rm mc} \nonumber \\
                   &- \langle s_x (s_{r} + s_{r-1} + \cdots + s_{-r+1} ) \rangle_{\rm mc} \nonumber \\
  &= \langle s_x ( - s_{r} + s_{-r} ) \rangle_{\rm mc} \, .
\end{align}
From the second to the third line, we used the translational invariance to shift from the index $x-1$ to $x$.
We consider the microcanonical ensemble with zero total stretch, momentum, and a fixed finite total energy.
We then impose the ensemble equivalence between the microcanonical ensemble with the ground canonical ensemble with the corresponding temperature $T$ and chemical potential $\mu$:
\begin{align}
  \langle s_x s_{r}  \rangle_{\rm mc} &={
                                       \int d s_1  \cdots d s_{N} \, s_x s_{r}\, 
                                       e^{-   ( V ({\bm s}) - \mu \sum_{i=1}^N s_{i} )/(k_{\rm B}T)}
                                     \over Z_{T,\mu}
                                       } \, , \\
  V({\bm s}) &= {1\over 2 \tN } \sum_{x=1}^{N} \sum_{r=1}^{N/2} {(s_x+s_{x+1}+\cdots +s_{x+r-1})^2 \over r^{\d}} \, \nonumber \\
  &= \sum_k |s_k|^2  {\sum_{r=1}^{N/2} {1\over \tN}  {\sin^2 (kr /2) \over r^{\d}}  \over 2 \sin^2 (k/2)} \, , 
\end{align}
where the Fourier transform is used in the last line. The boundary condition for this case is $s_{x+N}=s_x$.  Now, we take $\mu=0$, which corresponds to $\sum_{i=1}^N s_i=0$. With these parameters, we have
\begin{align}
h_r (k) &= {ik_{\rm B} T \over \sqrt{N}} {\sin (kr) \over \sum_{r=1}^{N/2} {1\over \tN} {\sin^2 (kr /2) \over r^{\d}}}   \, .
\end{align}
This is equivalent to Eq.(\ref{hrk1}) with $k_0=0$.

Note that here we do not prove the ensemble equivalence rigorously, but simply impose it. We remark that for short range interacting case, i.e., the original ME model, the ensemble equivalence can be rigorously proven \cite{basile2009cmp}.

\newpage 


\begin{thebibliography}{99}

\bibitem{campa2009physrep}
A. Campa, T. Dauxois, and S. Ruffo,
{\em Statistical mechanics and dynamics of solvable models with long-range interactions},
Phys. Rep. {\bf 480}, 57 (2009).

\bibitem{campa2014book}
A. Campa, T. Dauxois, D. Fanelli, and S. Ruffo,
{\em Physics of Long-Range Interacting Systems}
(Oxford University Press, Oxford, 2014).

\bibitem{thirring}
W. Thirring,
{\em System with Negative Specific Heat},
Z. Phys. {\bf 235}, 339 (1970).

\bibitem{hertel-thirring}
P. Hertel and W. Thirring,
{\em Free Energy of Gravitating Fermions},
Ann. Phys. (N.Y.) {\bf 63}, 520 (1971).

\bibitem{antoni-torcini-1998}
M. Antoni and A. Torcini,
{\em Anomalous diffusion as a signature of a collapsing phase in two-dimensional self-gravitating systems},
Phys. Rev. E {\bf 57}, R6233 (1998).

\bibitem{barre-mukamel-ruffo}
J. Barr\'{e}, D. Mukamel, and S. Ruffo,
{\em Inequivalence of Ensembles in a System with Long-Range Interactions},
Phys. Rev. Lett. {\bf 87}, 030601 (2001).

\bibitem{latora-rapisarda-tsallis}
V. Latora, A. Rapisarda, and C. Tsallis,
{\em Non-Gaussian equilibrium in a long-range Hamiltonian system},
Phys. Rev. E {\bf 64}, 056134 056134 (2001).

\bibitem{christodoulidi-tsallis-bountis}
H. Christodoulidi, C. Tsallis, and T. Bountis,
{\em Fermi-Pasta-Ulam model with long-range interactions: Dynamics and thermostatistics},
Europhys. Lett. {\bf 108}, 40006 (2014).

\bibitem{torcini-antoni-1999}
A. Torcini and M. Antoni,
{\em Equilibrium and dynamical properties of two-dimensional $N$-body systems with long-range attractive interactions},
Phys. Rev. E {\bf 59}, 2746 (1999).

\bibitem{latora-rapisarda-ruffo-1999}
V. Latora, A. Rapisarda, and S. Ruffo,
{\em Superdiffusion and Out-of-Equilibrium Chaotic Dynamics with Many Degrees of Freedoms},
Phys. Rev. Lett. {\bf 83}, 2104 (1999).

\bibitem{latora-rapisarda-ruffo-1998}
V. Latora, A. Rapisarda, and S. Ruffo,
{\em Lyapunov Instability and Finite Size Effects in a System with Long-Range Forces},
Phys. Rev. Lett. {\bf 80}, 692 (1998).

\bibitem{anteneodo-tsallis}
C. Anteneodo and C. Tsallis, 
{\em Breakdown of Exponential Sensitivity to Initial Conditions: Role of the Range of Interactions},
Phys. Rev. Lett. {\bf 80}, 5313 (1998).

\bibitem{bagchi-tsallis}
D. Bagchi and C. Tsallis,
{\em Sensitivity to initial conditions of a $d$-dimensional long-range-interacting quartic Fermi-Pasta-Ulam model: Universal scaling},
Phys. Rev. E {\bf 93}, 062213 (2016).

\bibitem{lepri2016book}
Edited by S. Lepri,
{\em Thermal Transport in Low Dimensions: From Statistical Physics to Nanoscale Heat Transfer},
Lecture Notes in Physics Vol. 921 (Springer-Verlag, Berlin, 2016). 

\bibitem{lepri2003physrep}
S. Lepri, R. Livi, and A. Politi,
{\em Thermal conduction in classical low-dimensional lattices},
Phys. Rep. {\bf 377}, 1 (2003).

\bibitem{dhar2008}
A. Dhar,
{\em Heat transport in low-dimensional systems},
Adv. Phys. {\bf 57}, 457 (2008).

\bibitem{jara2009}
M. Jara, T. Komorowski, and S. Olla,
{\em Limit Theorems for Additive Functionals of a Markov Chain},
Ann. Appl. Probab. {\bf 19}, 2270 (2009).

\bibitem{basile-olla-spohn-2010}
G. Basile, S. Olla, and H. Spohn,
{\em Energy Transport in Stochastically Perturbed Lattice Dynamics},
Arch. Ration. Mech. Anal. {\bf 195}, 171 (2010).

\bibitem{jara2015}
M. Jara, T. Komorowski, and S. Olla,
{\em Superdiffusion of Energy in a Chain of Harmonic Oscillators with Noise}
Commun. Math. Phys. {\bf 339}, 407 (2015).

\bibitem{olivares2016}
C. Olivares and C. Anteneodo,
{\em Role of the range of the interactions in thermal conduction},
Phys. Rev. E {\bf 94}, 042117 (2016).

\bibitem{bagchi2017xy}
D. Bagchi,
{\em Energy transport in the presence of long-range interactions},
Phys. Rev. E {\bf 96}, 042121 (2017).

\bibitem{iubini2018}
S. Iubini, P. Di Cintio, S. Lepri, R. Livi, and L. Casetti,
{\em Heat transport in oscillator chains with long-range interactions coupled to thermal reservoirs},
Phys. Rev. E {\bf 97}, 032102 (2018).

\bibitem{bagchi2017fpu}
D. Bagchi,
{\em Thermal transport in the Fermi-Pasta-Ulam model with long-range interactions},
Phys. Rev. E {\bf 95}, 032102 (2017).

\bibitem{xiong}
J. Wang, S. V. Dmitriev,and D. Xiong,
{\em Thermal transport in long-range interacting Fermi-Pasta-Ulam rings},
Phys. Rev. Research {\bf 2}, 013179 (2020).

\bibitem{miloshevich}
G. Miloshevich, J. P. Nguenang, T. Dauxois, R. Khomeriki, and S. Ruffo,
{\em Instabilities and relaxation to equilibrium in long-range oscillator chains},
Phys. Rev. E {\bf 91}, 032927 (2015).

\bibitem{basile2006prl}
G. Basile, C. Bernardin and S. Olla,
{\em Momentum Conserving Model with Anomalous Thermal Conductivity in Low Dimensional Systems},
Phys. Rev. Lett. {\bf 96}, 204303 (2006).

\bibitem{basile2009cmp}
G. Basile, C. Bernardin, and S. Olla,
{\em Thermal Conductivity for a Momentum Conservative Model},
Commun. Math. Phys. {\bf 287}, 67 (2009).


\bibitem{basile2016book}
G. Basile, C. Bernardin, M. Jara, T. Komorowski, and S. Olla,
in {\em Thermal Transport in Low Dimensions: From Statistical Physics to Nanoscale Heat Transfer},
edited by S. Lepri (Springer, Berlin, 2016).

\bibitem{lepri-monasterio-2009}
S. Lepri, C. Mej\'{i}a-Monasterio, and A. Politi,
{\em A stochastic model of anomalous heat transport: analytical solution of the steady state},
J. Phys. A: Math. Theor. {\bf 42}, 025001 (2009).

\bibitem{ft1} For the momentum conserving case, distribution should be defined for $(\{ {\bm s} \} , \{ {\bm p} \})$, for which the dynamics is derived by modifying the part of $\bA $.
  
\bibitem{protter} P. E. Protter, {\em Stochastic Integration and Differential
Equations}, (2nd Ed. Springer, 2005).

\bibitem{tamaki2017}
S. Tamaki, M. Sasada, and K. Saito,
{\em Heat Transport via Low-Dimensional Systems with Broken Time-Reversal Symmetry},
Phys. Rev. Lett. {\bf 119}, 110602 (2017).

\bibitem{saito-sasada-2017}
K. Saito and M. Sasada,
{\em Thermal conductivity for a system of harmonic oscillators with noise in a magnetic field},
M. Commun. Math. Phys. {\bf 361}, 951 (2018).

\bibitem{kth}
  R. Kubo, M. Toda, and N. Hashitsume,
  {\em Statistical Physics II: Nonequilibrium Statistical Mechanics}, (Springer, 2012)

\bibitem{saitodhar}
K. Saito and A. Dhar, {\em Heat conduction in a three dimensional anharmonic crystal}, 
Phys. Rev. Lett. {\bf 104}, 040601 (2010).


\bibitem{pre2010}
D. Xiong, J. Wang, Y. Zhang, and H. Zhao, {\em Heat conduction in two-dimensional disk models}, Phys. Rev. E {\bf 82}, 030101(R) (2010).


\bibitem{beijeren2012}
H. van Beijeren,
{\em Exact Results for Anomalous Transport in One-Dimensional Hamiltonian Systems},
Phys. Rev. Lett. {\bf 108}, 180601 (2012).

\bibitem{spohn2014}
H. Spohn,
{\em Nonlinear Fluctuating Hydrodynamics for Anharmonic Chains},
J. Stat. Phys. {\bf 154}, 1191 (2014).



\bibitem{das2014}
S. G. Das, A. Dhar, K. Saito, C. B. Mendl, and H. Spohn,
{\em Numerical test of hydrodynamics fluctuation theory in the Fermi-Pasta-Ulam chain},
Phys. Rev. E {\bf 90}, 012124 (2014).





\end{thebibliography}

\end{document}